\documentstyle[12pt,epsf]{article}
\newcommand{\bfig}{\begin{figure}}
\newcommand{\efig}{\end{figure}}

\def\ba{\begin{array}}
\def\ea{\end{array}}
\def\bea{\begin{eqnarray}}
\def\eea{\end{eqnarray}}

\newcommand{\trans}[1]{{t\  \atop \displaystyle \:}\!\!\!\! {#1}}

\def\one-loop{\mbox{\scriptsize one-loop}}

\def\s{\sigma}

\def\psivector{\mbox{\boldmath$\psi_{Q}$}}

\def\G{\Gamma}

\catcode`\@=11

\@addtoreset{equation}{section}
\def\theequation{\arabic{section}.\arabic{equation}}

\def\@normalsize{\@setsize\normalsize{15pt}\xiipt\@xiipt
\abovedisplayskip 14pt plus3pt minus3pt%
\belowdisplayskip \abovedisplayskip
\abovedisplayshortskip  \z@ plus3pt%
\belowdisplayshortskip  7pt plus3.5pt minus0pt}

\def\small{\@setsize\small{13.6pt}\xipt\@xipt
\abovedisplayskip 13pt plus3pt minus3pt%
\belowdisplayskip \abovedisplayskip
\abovedisplayshortskip  \z@ plus3pt%
\belowdisplayshortskip  7pt plus3.5pt minus0pt
\def\@listi{\parsep 4.5pt plus 2pt minus 1pt
            \itemsep \parsep
            \topsep 9pt plus 3pt minus 3pt}}

\def\underline#1{\relax\ifmmode\@@underline#1\else
        $\@@underline{\hbox{#1}}$\relax\fi}
\@twosidetrue
\relax

\catcode`@=12

\catcode`\@=11

\def\section{\@startsection{section}{1}{\z@}{3.5ex plus 1ex minus
   .2ex}{2.3ex plus .2ex}{\large\bf}}

\def\thesection{\Roman{section}.}

\def\appendix{\setcounter{section}{0}
        \def\thesection{Appendix }
        \def\theequation{\Alph{section}.\arabic{equation}}}
\def\ps@headings{\def\@oddfoot{}\def\@evenfoot{}
\def\@oddhead{\hbox{}\hfill
        \makebox[.5\textwidth]{\raggedright\ignorespaces --\thepage{}--
        \hfill {}}}
\def\@oddhead{\hbox{}\hfill --\thepage{}-- \hfill
        {}}
\def\@evenhead{\@oddhead}
\def\subsectionmark##1{\markboth{##1}{}}
}

\ps@headings

\catcode`\@=12

\relax

\def\figcap{\section*{Figure Captions\markboth
        {FIGURECAPTIONS}{FIGURECAPTIONS}}\list
        {Fig. \arabic{enumi}:\hfill}{\settowidth\labelwidth{Fig. 999:}
        \leftmargin\labelwidth
        \advance\leftmargin\labelsep\usecounter{enumi}}}
 \relax
\def\tablecap{\section*{Table Captions\markboth
        {TABLECAPTIONS}{TABLECAPTIONS}}\list
        {Table \arabic{enumi}:\hfill}{\settowidth\labelwidth{Table 999:}
        \leftmargin\labelwidth
        \advance\leftmargin\labelsep\usecounter{enumi}}}
 \relax
\def\reflist{\section*{References\markboth
        {REFLIST}{REFLIST}}\list
        {[\arabic{enumi}]\hfill}{\settowidth\labelwidth{[999]}
        \leftmargin\labelwidth
        \advance\leftmargin\labelsep\usecounter{enumi}}}
 \relax

\catcode`\@=11

\def\ps@headings{\def\@oddfoot{}\def\@evenfoot{}
\def\@oddhead{\hbox{}\hfill
        \makebox[.5\textwidth]{\raggedright\ignorespaces --\thepage{}--
        \hfill {}}}
\def\@evenhead{\@oddhead}
\def\subsectionmark##1{\markboth{##1}{}}
}

\ps@headings

\relax

\newskip\humongous \humongous=0pt plus 1000pt minus 1000pt

\newif\ifdtup

\def\beq{\begin{equation}}
\def\eeq{\end{equation}}

\def\beqn{\begin{eqnarray}}
\def\eeqn{\end{eqnarray}}
\relax

\def\G2{{\; \rm GeV/}c^2}
\def\G{\; \rm GeV}

\def\dotx{\dotx{\dot\overline{x}}}

\relax

\begin{document}
\begin{titlepage}
\begin{flushright}
       {\normalsize OU-HET 320 \\  hep-th/9909075\\
         September, 1999 }
\end{flushright}
%\vfill
%
\begin{center}
  {\large \bf Nonabelian Monopoles from Matrices: \\
      Seeds of the Spacetime Structure }
\footnote{This work is supported in part
 by the Grant-in-Aid  for Scientific Research (10640268)
 and Grant-in-Aid  for Scientific Research fund (97319)
from the Ministry of Education, Science and Culture, Japan.}
%\end{center}

%\begin{center}
\vfill
         {\bf B.~Chen, ~H.~Itoyama}  \\
            and \\
              {\bf H.~Kihara}\\
%\vfill
        Department of Physics,\\
        Graduate School of Science, Osaka University,\\
        Toyonaka, Osaka 560-0043, Japan\\
\end{center}
\vfill
%%%%%%%%%%%%%%%%%%%%%%%%%%%%%%%%%%%%%%%%%%%%%%%%%%%%%%%%%%%%%
%%%%%%%%%%%%%%%%%%%%%%%%%%%%%%%%%%%%%%%%%%%%%%%%%%%%%%%%%%%%%
%\newpage
\begin{abstract}
  We study the expectation value of (the product) of the
 one-particle projector(s) in the reduced matrix model and matrix
 quantum mechanics  in general.  This quantity is given by
 the nonabelian Berry phase: we discuss
  the relevance of this with regard to the spacetime structure.
 The case of the $USp$ matrix model is examined from this respect.
 Generalizing our previous work, we carry out the complete
 computation  of this quantity which takes into account both the
 nature of the degeneracy of the fermions and the presence of
 the spacetime points belonging to the antisymmetric
 representation.
We find the singularities as those of the $SU(2)$ Yang monopole
 connection as well as the pointlike singularities in $9+1$ dimensions
  coming from its  $SU(8)$ generalization.  The former type of singularities,
 which extend to four of the directions lying in the
 antisymmetric representations,  may be regarded as seeds of our four
 dimensional spacetime structure and is not shared by the $IIB$ matrix model.
  From a mathematical viewpoint, these connections can be generalizable
  to arbitrary odd space dimensions due to the nontrivial nature of the
 eigenbundle and the Clifford module structure.

\end{abstract}
\vfill
\end{titlepage}
%%%%%%%%%%%%%%%%%%%%%%%%%%%%%%%%%%%%%%%%%%%%%%%%%%%%%%%%%%%%%
%%%%%%%%%%%%%%%%%%%%%%%%%%%%%%%%%%%%%%%%%%%%%%%%%%%%%%%%%%%%%

\section{Introduction}

Continuuing studies in matrix models for superstrings and M theory
 \cite{BFSS, IKKT, DVV, IT1, IT2} indicate
  that we are in a stage of obtaining a renewed understanding of
 old  notions such as compactification and spacetime distribution in
  this constructive  framework. These physical quantities 
 are obtained after  integrations of matrices and are no longer fixed
 input parameters  or backgrounds.
 Another feature common to these models is that the actions contain
 terms bilinear in fermions.  This is, of course, related
 to the D brane in the RR sector as the absence of these bilinears imply
  the absence of the RR sector of the model \footnote{ The reduced matrix
 model provides a constructive definition to
 the Green-Schwarz superstrings in the Schild gauge.
  A subsector of the state space which contains
 fermion bilinears is able to see the RR sector of the superstrings. }.

   The major objective of this paper is to enlighten the spacetime structure
  and the presence of solitonic objects revealed by the fermionic
 integrations of the matrix models.
  These are represented by the behavior of the spacetime points, ( or $D0$ branes
 in \cite{BFSS}), 
 which are the eigenvalue distributions or the diagonal elements of the bosonic
  matrices.  The effective dynamics of the spacetime points 
 is obtained by carrying out the integrations of the remaining degrees of freedom:
   we will carry out  the half of them  represented by the fermions.
  Our interest, therefore, lies in a collection of individual fermionic
 eigenmodes obtained from the
 fermionic part ( denoted generically by $S_{fermion}$) of the action upon
 diagonalization.
  An object which, we find, plays  a role of revealing singularities
   as those of the bosonic parameters ( in particular, those of
 the spacetime points)  is an expectation
 value of the one-particle projector belonging to each of the fermionic
 eigenmodes.  (See eq. (\ref{eq:objecttext})).

  We see that this expectation value of the projector is generically given by
  the nonabelian Berry phase \cite{Berry, WZ}.
 In matrix models of superstrings and M theory, this result offers
 spacetime interpretation:  this is because the parameter space, where
 the connection one-form lives, is that of the spacetime points or $D0$ branes.
 An interesting case that we study as our major example is the $USp$ reduced
 matrix model \cite{IT1, IT2, IM, CIK, Itsu, ItsuNishi}.  We will carry out
 an explicit evaluation of the nonabelian Berry phase factor for different types of the projectors.
 Our computation leads us to the $su(2)$ Lie algebra valued connection one-form
  known as the Yang monopole \cite{Yang} in five spatial dimensions and 
 its $su(8)$ generalization in nine spatial dimensions.
  This latter one, to the best of our knowledge, has not appeared
  in physics context before. The existence of the nontrivial eigenbundles based
 on the first quantized hamiltonian with gamma matrices and the orthogonal
 projection operators ensures that these  nonabelian connections are
 straightforwardly generalized to arbitrary odd spatial dimensions.

  The conclusion derived from our computation in the context of
  the $USp$ matrix model is that there exist
   singularities extending to four of the directions of the spacetime points
 which lie in the antisymmetric representation. 
 These singularities are represented by the Yang monopole. 
  In addition, we find pointlike singularities in $9+1$ dimensions
  which are represented by its $SU(8)$ generalization. The former
  type of singularities may be regarded as being responsible for our four
 dimensional spacetime structure and is not shared by the $IIB$ matrix model
 \footnote{ The signature of the four directions which
  the Yang monopole is extended to is euclidean in the model.
  We have nothing to say on how to make the signature Minkowskiian
  and on the extension of the spacetime in the $v_{0}$ direction.}.
  It is noteworthy that the requirement of having $8+8$ supersymmetries
 brings us such possibility.

  Before moving to the next section, let us  mention the procedure
  to formulate  the expectation value of the one-particle projector
  in the reduced matrix models. (See eq. (\ref{eq:objecttext})).
  To say things short, these are obtained from short time/infinite
  temperature limit of the matrix quantum mechanical system,
  where a path in the parameter space can be readily introduced.
    This is equivalent to imposing a periodicity  with period $R$ on one of the
 original bosonic matrices, say, $v_{0}$ and to letting this period
 go to infinity in the end. Machinery to deal with these situations has been
 developed in \cite{WT}.  We mention here the calculation of
 \cite{nekra, kos}, which
 is similar in spirit to ours.   There the exact computation of the
 partition function of the $IIB$ matrix model \cite{IKKT}  has been
 carried out as the limit of infinite temperature 
  of the path integral of the BFSS \cite{BFSS} quantum mechanical model.
  This is the same limiting procedure as ours although
  we will measure the one-particle projector instead of unity.
 (See eq.~(\ref{eq:opex}). )

 In the next section, we describe the formulation and the procedure of our
 computation indicated above  after a brief review on the $USp(2k)$ matrix
 model.
In section III, we decompose the fermionic part of
the action into the bases of the adjoint, antisymmetric and fundamental
representations appropriate to our computation.
 This amounts to diagonalizing it  for the case
    that the bosonic matrices  are diagonal. 
Unlike our previous studies \cite{IM, CIK}, we do not set
 the spacetime points lying in the antisymmetric representation to zero.
 This turns out to improve substantially
 our picture of the spacetime formation suggested by the model.
In section IV, we compute the nonabelian Berry phase
for three types of actions obtained in section III.
The connection one-forms we find are the nonabelian
$SU(2)$ Yang monopole in five dimensions and
  its $SU(8)$ generalization to nine dimensions.
  The degenerate state space originating from  the spinorial space is
 responsible for
  making these nonabelian gauge fields. In section V, we summarize the
 spacetime picture emerging from our computation.  
 In section VI, we discuss the generalization of the Yang monopole
 in arbitrary odd space dimensions by clarifying  the eigenbundle
 structure associated with the nonabelian Berry connection.
  Detail of the basis decomposition in section III
  is collected in the Appendix.
  Some of the papers on fermionic and bosonic
 integrations of matrix models are listed in \cite{fermi}.

\section{ Nonabelian Berry Phase and Matrix Models }
 The goal of this section  is to establish that
  the expectation value of the one-particle projector
   of a fermionic eigenmode is given by the nonabelian Berry phase.
  This is true in general, in particular, in matrix models
 (both matrix quantum mechanics and reduced matrix models)
  containing fermion bilinears. Besides the examples we discuss below,
our discussion here will apply to a variety of models obtained, for instance, 
 by a truncation from supersymmetric field theories in various
 dimensions \cite{sq}.  
As we will occasionally  refer to the case of the $USp(2k)$
 matrix model--  the major example  of our paper-- already in this section,
  we will present the brief review of this model in the first subsection
 and  defer the major discussion to the second subsection.

\subsection{some preliminaries}
  To begin with, the $usp$ Lie algebra is defined by
\beq
\label{eq:lie}
\mbox{usp}(2k) \equiv \{A \in \mbox{u}(2k)|\: ^tAF + FA=0 \} \;\;,
\eeq
 and the antisymmetirc representation is  defined by
\beq
\label{eq:asym}
\mbox{asym}(2k) \equiv \{A \in \mbox{u}(2k)|\: ^tAF - FA=0 \} \;\;.
\eeq
 Here $F$ is an antisymmetric matrix with nonzero determinant
  and can be chosen  as
\beq
\label{eq:F}
 F = \left(\begin{array}{cc}0&-{\bf 1}_{k}\\{\bf 1}_{k} &0\end{array}\right)
 \;\;\;.
\eeq
 It is easy to recognize
\beq
\mbox{u}(2k) = \mbox{usp}(2k) \oplus \mbox{asym}(2k)\;\;.
\eeq
  A representaion of eq.~(\ref{eq:lie}) in accordance with the choice 
 ( eq.~(\ref{eq:F}) )  is
\begin{eqnarray}
A &\equiv& \left( 
\begin{array}{cc}
  H  &B\\
\overline{B}&-\overline{H}
\end{array}\right)   \in  \mbox{usp}(2k) \;\;\;,   \\
H^{\dag} &=& H\;\;, \;\;
\trans{B} = B\;\;,  \nonumber
\end{eqnarray}
  while  that  of eq. (\ref{eq:asym}) is
\begin{eqnarray}
A &\equiv& \left( 
\begin{array}{cc}
H&B\\
-\overline{B}&\overline{H}
\end{array}\right)  \in \mbox{asym}(2k)  \;\;\;,  \\
H^{\dag} &=& H \;\;, \;\;
\trans{B} = -B \;\;.  \nonumber
\end{eqnarray}

 Let us recall here some aspects of the reduced $USp(2k)$ matrix model which
 are relevant to our discussion in what follows.
 The definition, the criteria and the rationale leading to the model
 as descending from  $typeI$ the superstrings are
 ellaborated fully in ref.\cite{IT1,IT2}.  We will therefore not repeat
 these here.

 Let $\hat{\Omega}$ be a projection operator acting on $2k \times 2k$ hermitean
 matrices.
For the ten bosonic matrices, $\hat{\Omega}$   projects the $0,1,2,3,4,7$ 
 components onto  the adjoint representation and the $5,6,8,9$ components
  onto the antisymmetric representation:
\begin{eqnarray}
v_M &=& (v_{\mu} , v_n)\;\;,   \nonumber  \\
v_{\mu} &\in& \mbox{usp}(2k) ,\:\:\:\:\:\mu = 0,1,2,3,4,7\;\;\;,   \\
 v_{n} &\in& \mbox{asym}(2k)\:\:\:\:\:\:n=5,6,8,9 \;\;\;.  \nonumber
\end{eqnarray}
 As for the fermions,  $\hat{\Omega}$ splits the thirty two component
 Majorana-Weyl spinor (sixteen real degrees of freedom) in $9+1$ dimensions
 into an eight component spinor belonging to the adjoint representation and
  another eight component spinor belonging to the antisymmetric representation:
\begin{equation}
\label{eq:psi}
\Psi = \Psi_{adj} + \Psi_{asym}\;\;,
\end{equation}
 where
\begin{eqnarray}
\label{eq:adjanti}
\Psi_{adj} &\equiv& \trans{(\lambda_1 ,0,\lambda_2,0,0,0,0,0,0,
\overline{\lambda_1},0,\overline{\lambda}_2,0,0,0,0)}\;\;\;, \nonumber \\
\Psi_{asym} &\equiv& \trans{(0,0,0,0,\psi_1,0,\psi_2,0,0,0,0,0,0,
\overline{\psi}_1,0,\overline{\psi}_2)}\;\;\;.
\end{eqnarray}
  Modulo labelling the indices, these projections are
  determined by the requirement of having $8+8$ supersymmetries.
  Finally, we add degrees of freedom  correponding to an open string degrees
 of freedom  while preserving supersymmetry. This amounts to adding
  $n_{f}=16$ of the hypermultiplets in the fundamental representation in
 the $4d$ language.
  We display these  degrees of freedom  by
  the complex $2n_{f}$ dimensional vectors ( see the appendix A of
 \cite{Itsu})
\beqn
 {\bf Q} \equiv  \left\{ \begin{array}{ll}
       Q_{(f)}\;\;, &  f=1 \sim n_{f}  \\
     F^{-1} \tilde{Q}_{(f- n_{f})}\;\;, & f=n_{f} +1 \sim 2n_{f} \;,
   \end{array}
   \right.   \;\;\;
 {\bf Q}^{*} \equiv  \left\{ \begin{array}{ll}
       Q_{(f)}^{*} \;\;, &  f=1 \sim n_{f}  \\
      \tilde{Q}_{(f- n_{f})}^{\ast} F \;\;, & f=n_{f} +1 \sim 2n_{f}  \;.
   \end{array}
   \right.      
\eeqn
\beqn
  \psivector   \equiv  \left\{ \begin{array}{ll}
      \psi_{ Q_{(f)}} \;\;, &  f=1 \sim n_{f}  \\
     F^{-1} \psi_{\tilde{Q}_{(f- n_{f})} } \;\;, & f=n_{f} +1 \sim 2n_{f} \;, 
   \end{array}
   \right.     
   \psivector^{\ast}    \equiv  \left\{ \begin{array}{ll}
       \overline{\psi}_{Q_{(f)} } \;\;, &  f=1 \sim n_{f}  \\
    \overline{\psi}_{\tilde{Q}_{(f- n_{f})} } F \;\;, & f=n_{f} +1
 \sim 2n_{f} \;.
   \end{array}
   \right.    
\eeqn

 Let us turn to the action of the model.  It is represented as
\beq
S_{USp}= S_{closed} + \Delta S \;\;\;,
\eeq
  where
\begin{eqnarray}
S_{closed} = \frac{1}{g^2} \mbox{Tr} \left\{ \frac{1}{4} \left[ v_M ,
 v_N \right] \left[ v^M , v^N \right] 
- \frac{1}{2} \overline{\Psi} \Gamma^M \left[ v_M , \Psi \right] \right\}  
\end{eqnarray}
 is the closed string sector of the model.  We denote the fermionic part by
\beq
S_{MW} \equiv -\frac{1}{2g^2} \mbox{Tr} \left(
 \overline{\Psi} \Gamma^M \left[ v_M , \Psi \right] \right) \;\;. 
\eeq
  The remainder of the action
\begin{eqnarray}
\Delta S  =  \left\{ {  S_{g-s} + {\cal V}_{scalar} + S_{mass}} +
 {  S_{g-f} + S_{Yukawa}}\right\}  \;\;,
\end{eqnarray}
  consists of the parts which depend on the fundamental hypermultiplet.
  We only spell out the parts relevent to our subsequent discussion.
\beqn
  S_{g-f} &=& \frac{1}{g^{2}} \left(  \psivector^{\ast}
  \overline{\sigma}^{m} v_{m} \cdot \psivector
 + i \sqrt{2} {\bf Q}^{*} \lambda \cdot
  \psivector  - i \sqrt{2} \psivector^{\ast}
 \overline{\lambda} \cdot {\bf Q}   \right) \;,  \label{eq:gf}  \\
  S_{Yukawa} &=& - \frac{1}{g^2}\left( \sum_{(c_{1}, c_{2})= 
  (Q, \tilde{Q}), (Q, \Phi_{1}), (\Phi_{1},\tilde{Q})}
 \frac{\partial^{2} W_{matter}}
{\partial C_{1} \partial C_{2}} \psi_{C_{2}} \psi_{C_{1}} + h.c. \right) 
  \nonumber \\
 &=& \frac{1}{g^2}\left( \frac{1}{2}   \psivector  \cdot  \Sigma   F
  \left( \sqrt{2} \Phi_{1} + M \right) \psivector
 + \sqrt{2} {\bf Q} \cdot  \Sigma  F \psi_{\Phi_{1}}  \psivector
  +  h.c. \right) \;\;. \label{eq:Yukawa}
\eeqn
  Here
\beqn
  \Sigma  &\equiv&  \left(
 \begin{array}{cc}
   0 & I  \\
   I & 0  
  \end{array}   \right)\;\;,  \\
M &\equiv&  {\rm diag} \left( m_{(1)}, \cdots, m_{(n_f)}
 - m_{(1)}, \cdots, -m_{(n_f)} \right) \;\;,  \\
  W_{{\rm matter}} &=&  \sum_{f=1}^{n_{f}} \left( m_{(f)}
 \tilde{Q}_{(f)} Q_{(f)}+ \sqrt{2} \tilde{Q}_{(f)} \Phi Q_{(f)} \right)\;\;,
\eeqn
  and $\cdot$ implies the standard inner product with respect to the
  $2n_{f}$ flavour indices.
  For a more complete discussion, see \cite{Itsu}.

\subsection{one-particle projector and nonabelian Berry phase}

  Let us first imagine diagonalizing some action $S_{fermion}$
 which is bilinear in fermions. In the example of the last subsection,
  this is given by the fermionic part of the  action
\beqn
\label{eq:fermiS}
  S_{fermion} \equiv  S_{MW} + S_{gf}+ S_{Yukawa} \;\;. 
\eeqn
  In general, $S_{fermion}$ can be written as
\beqn
\label{eq:diagf}
  S_{fermion} =   \sum_{\alpha_{\ell}} \sum_{\ell} \lambda_{\ell}
     \bar{\xi_{\ell}}^{\alpha_{\ell}}  \xi_{\ell}^{\alpha_{\ell}} \;\;\;.
\eeqn
  Here $\xi_{\ell}^{\alpha_{\ell}}$ is the 
   fermionic eigenmode belonging to an eigenvalue $\lambda_{\ell}$ and 
 $\alpha_{\ell}$ labels its degeneracy.
  As mentioned in the introduction, we would like to
 evaluate an expectation value of the one-particle projector
\beq
\label{eq:calP}
\hat{{\cal P}}_{\ell}^{\alpha \alpha^{\prime}}  \equiv 
\xi_{\ell}^{\dagger \alpha} \mid \Omega  \rangle \langle \Omega \mid 
\xi_{\ell}^{\alpha^{\prime}} \;\;,
\eeq
  with $\Omega$ being the Clifford vacuum  which annihilates  half
  of  the fermions
\beq 
 \left( {\bf b}^{A},  \bar{ {\bf {b}} }_{A} \right) \;\;,
\eeq 
 which are   $ \left( \Psi, \bar{\Psi}, 
\mbox{\boldmath $\psi$}_{\bf Q} \mbox{\boldmath $\psi$}^{*}_{\bf Q} \right)$
  in the model of the last subsection.
  The eigenmodes $\xi_{\ell}^{\alpha_{\ell}}$ can be written as
\beqn
\label{eq:bpsi}
   \xi_{\ell}^{\alpha} = \sum_{A} b^{A}\psi_{\ell A}^{\alpha}\;\;.
\eeqn

 The expectation value is defined through the integrations over
  the fermionic variables of the model.  A formula  that we find in the end
   (eq.~(\ref{eq:opex}) ) and  the one we use (eq.~(\ref{eq:formulaPB})  )
  in the subsequent sections  are obtained from the short time/infinite
 temperature limit of the corresponding quantum mechanics,
    in which the path dependence  can be easily introduced.
 In order to argue more directly that  this quantity can be defined in the
  reduced models, we start with imposing
a periodicity constraint on one of the directions, say, $v_{0}$:
\beqn
\label{eq:period}
  {\cal S} v_{M} {\cal S}^{-1} &=& v_{M} +  R {\bf 1} \delta_{M.0}\;\;.
\eeqn
The size of the matrices is necessarily infinite dimensional in order to
permit solutions to eq.~(\ref{eq:period}). Each matrix divides into an infinite
number of blocks. The shift operator ${\cal S}$ acts on each block  and moves
 it diagonally by one in our situation. 

  Let us introduce 
\beqn
\label{eq:objecttext}
     \langle \langle \;\;
 \hat{{\cal P}}_{\ell}^{\alpha \alpha^{\prime} }  \;\; \rangle \rangle_{\Gamma}
  \equiv  \lim_{R \rightarrow \infty} \int \left[ {\cal D} {\cal \kappa}
 \right]\left[ {\cal D} {\cal Z} \right]
    e^{ i  S( {\cal \kappa}, {\cal Z};  {\cal Z}^{\prime},
  {\cal Z}^{\prime \prime}) }   
  \langle {\cal Z}^{\prime} \mid   \hat{{\cal P}}_{\ell}^{\alpha
 \alpha^{\prime} }  \mid  {\cal Z}^{\prime \prime}  \rangle  \;\;.
\eeqn
  Here  $S( {\cal \kappa}, {\cal Z};  {\cal Z}^{\prime},
  {\cal Z}^{\prime \prime})$  and  
$\langle {\cal Z}^{\prime} \mid   \hat{{\cal P}}_{\ell}^{   }
  \mid  {\cal Z}^{\prime}  \rangle$ are the Grassmann coordinate
 representation of $S_{fermion}$ and that of
 $\hat{{\cal P}}_{\ell}^{\alpha \alpha^{\prime}}$ respectively. We have
 indicated the end point constraints and the path $\Gamma$ in
 eq.~(\ref{eq:objecttext}).  These will become clearer shortly. 
   We  will also consider
\beq
\label{eq:prodob}
   \langle \langle \;\;
  \prod_{\ell \in {\cal I} }
 \hat{{\cal P}}_{\ell}^{\alpha \alpha^{\prime}}\;\; \rangle \rangle_{\Gamma}
  \;\;,
\eeq
  where ${\cal I}$ is a subset of all eigenmodes and  the case of our interest
 is the one in which this subset is  over the
 eigenmodes belonging to  the positive eigenvalues.
  This choice is motivated by the Dirac sea filling.

  In principle,  one can diagonalize eq.~(\ref{eq:fermiS})  for general
  $v_{M}$ and ${\bf Q}$. We will, however, restrict ourselves to the case
\beq
 v_{M} = X_{M} = {\rm diagonal}\;\;, \;\;\; {\bf Q}=0 \;\;.
\eeq
  Explicit diagonalization  of  eq.~(\ref{eq:fermiS}) to the form of
 eq.~(\ref{eq:diagf}) in this case will be carried out in the next section.
  
  Let us  convert eq.~(\ref{eq:objecttext})
  into the Fourier transformed variables and this helps us understand
   the limiting procedure in eq.~(\ref{eq:objecttext}) better.
\beqn
\label{eq:pathint}
    \langle \langle \;\;
 \hat{{\cal P}}_{\ell}^{\alpha \alpha^{\prime}}  \;\; \rangle \rangle_{\Gamma}
  &=&  \lim_{\beta \rightarrow 0}  \int  dz^{\prime} dz^{\prime \prime}
 {\cal F} \left( z^{\prime}; \beta  \mid z^{\prime \prime} ; 0 \right)_{\Gamma}
  \langle z^{\prime} \mid   \hat{{\cal P}}_{\ell}^{\alpha,\alpha^{\prime} }
  \mid  z^{\prime \prime}  \rangle    \;\;,  \\
{\cal  F} &=&  \int  \left[ {\cal D} \kappa \left( \cdot \right)
 \right]\left[ {\cal D} z \left( \cdot \right) \right]
    e^{  i   \int_{0}^{\beta} d \beta^{\prime}
   {\cal L}_{fermion} \left( \kappa \left( \beta^{\prime} \right), 
  z \left( \beta^{\prime} \right) ;   z^{\prime}, z^{\prime \prime},
  X_{M}= X_{M}(\beta^{\prime}) \right) }  \;\;.
\eeqn
  Here 
\beq
  \beta = 2\pi/ R \;\;.
\eeq
  and
  ${\cal L} \left( \cdots  \right) $ is 
  the Grassman coordinate representation of the matrix quantum mechanics
 Lagrangian  which is obtained from $S_{fermi}$
  by the susbstitution \footnote{ An infinite normalization $\delta(0)$
 is involved in the relation $S_{fermi} \sim{\cal L}_{fermion}$, which
  is related to the problem of the scaling limit. We will  simply absorb
  this in the coupling $g^{2}$.}
\beqn
  v_{0} \rightarrow   i \frac{d}{dt} \;\;.
\eeqn
  Here  we have chosen the $v_{0}=0$ gauge.  In  the operator representation
  with corresponding Hamiltonian  denoted by 
\beq
  H \left(\beta^{\prime}\right)  \equiv
H  \left[ {\bf b}^{A}, \; \bar{ {\bf b} }^{A}, \;  
  \mid \Gamma ; \;  X_{M}= X_{M}(\beta^{\prime}) \right] \;\;\;,
\eeq
 Eq.~(\ref{eq:pathint}) is  written as
\beqn
\label{eq:opex}
    \langle \langle \;\; \hat{{\cal P}}_{\ell}^{\alpha \alpha^{\prime}}
  \;\; \rangle \rangle_{\Gamma} 
= \lim_{\beta \rightarrow 0}   Tr_{fermion}  \left(  \left(-\right)^{F}
    e^{ - i   \int_{0}^{\beta} d \beta^{\prime} H \left(\beta^{\prime}\right) }
  {\cal P}_{\ell}^{\alpha \alpha^{\prime} }   \right)   \;\;\;. 
\eeqn
  Reducing this expression  into that of the first quantized quantum
 mechanics, we find  that this  quantity is nothing but the time evolution
  of the $\ell$-th  degenerate eigenfunction $\psi_{\ell}^{\alpha}$.
  ( Note that this $\psi_{\ell}^{\alpha}$ is the same as the one  appearing
  in the original expression eq.~(\ref{eq:bpsi}). )
 The generic expression is known to consist of the energy dependent
 dynamical phase and the nonabelian Berry phase \cite{Berry,WZ}. 
This latter phase factor
 is given by the path-ordered exponential of the loop integral of the
  connection one-form ${\cal A}_{\ell} \left( X_{M} \right)$.   We obtain
\beqn
 \lim_{\beta \rightarrow 0}  \left( P \exp \left[ -i 
 \int_{0}^{\beta} d \beta^{\prime}
  E_{\ell} \left(   X_{M}(\beta^{\prime}) \right)
     -i  \oint_{\Gamma} {\cal A}_{\ell} \left( X_{M} \right)
    \right] \right)^{\alpha  \alpha^{\prime}} \;\;.
\eeqn
  Here 
\beq
{\cal A}_{\ell} \left( X_{M} \right)  = -i 
 \psi_{\ell}^{\dagger} d \psi_{\ell}
  = -i \sum_{A}   \psi_{\ell A}^{ \alpha \dagger}
 d \psi_{\ell A}^{\alpha^{\prime}} \;\;. 
\eeq
  and the nonabelian gauge field ${\cal A}_{\ell} \left( X_{M} \right)$
 originates from the degenerate eigenfunction.

  Finally letting $R$ large or the time period  $\beta$  short,
 we find
\beqn
\label{eq:formulaPB}
     \langle \langle \;\;
 \hat{{\cal P}}_{\ell}^{\alpha \alpha^{\prime}}\;\; \rangle \rangle_{\Gamma}
  =
 P \exp \left[  -i  \oint_{\Gamma} {\cal A}_{\ell} \left( x_{M} \right)
    \right]^{\alpha \alpha^{\prime}} \;\;,
\eeqn
  separating the phase of topological origin from the dynamical phase
\footnote{  With regard to  the last footnote,  we have adopted  here the
 normalization of  the energies/eigenvalues as quantum mechanics.}.

\bigskip

\section{Complete decomposition of the fermionic part of the action}

In this section, we will show how to decompose the fermionic part of the
 action. The diagonal part  of the fermions has no contribution to the action
  in the present treatment.
 We only need to focus on the
 off-diagonal part. After expanding by the bases of Lie algebra generators
 of $USp(2k)$, the fermionic part of the action finally can be classified
 into three types.

\subsection{the case of the $IIB$ matrix model}
 Let us consider  
the fermionic part  of the action of $IIB$ reduced matrix model:
\begin{equation}
{\cal S} = \frac{1}{2}\mbox{Tr}\overline{\Psi} \Gamma^{M}
 \left[ X_{M} , \Psi \right] \;\;.
\end{equation}
 The matrices are all $u(2k)$ Lie algebra valued. 
Here, for brevity, we ignore  the coupling and the minus sign in the action.
  These will be put back in the next section.
\beq
X_M=\left(\ba{ccc}
x^1_M& &0\\
&\ddots& \\
0& &x^N_M
\ea \right).
\eeq

Using the set of bases defined by 
\begin{equation}
(E_{ab})_{ij} \equiv \delta_{ai} \delta_{bj} \;\;,
\end{equation}
  we can write the bases of the Lie algebra generators of $U(N)$, which are
 Hermitian matrices,  as
\begin{eqnarray}
H_a &=& E_{a,a} - E_{a+1,a+1}\:\:\:\:( a=1,\cdots,N-1) \;\;,\\
S_{a,b} &=& E_{a,b} + E_{b,a}\:\:\:\:( a < b ) \;\;,\\
T_{a,b} &=& -i ( E_{a,b} - E_{b,a})\:\:\:\:( a < b ) \;\;.
\end{eqnarray}
  Here
\begin{eqnarray}
H_a = \left( 
\begin{array}{cccc}
 & & &\\
 &1 & &\\
&  &-1 &\\
& & & 
\end{array}
 \right) \;\;, \;\;\;
S_{a,b}= \left( 
\begin{array}{cccc}
 & & &\\
 & &1 &\\
&1 & &\\
& & & 
\end{array}
 \right) \;\;, \;\;\;
T_{a,b} = \left( 
\begin{array}{cccc}
 & & &\\
 & & -i&\\
& i& &\\
& & & 
\end{array}
 \right)  \;\;\;.
\end{eqnarray}

Let us decompose $\Psi$ into the diagonal part $\psi$ and the off-diagonal
 part $\chi$:
\begin{equation}
\Psi = \psi + \chi \;\;.
\end{equation} 
  The component expansion  of $\chi$ reads
\begin{equation}
\chi = \sum_{a<b} \left( \chi_S^{ab} S_{ab} + 
\chi_T^{ab} T_{ab} \right) \;\;\;.
\end{equation}

Now the action depends only on $\chi_S$ and $\chi_T$:
\begin{eqnarray}
{\cal S} = i \sum_{a<b} \left\{ - \overline{\chi}_S^{ab}
 \Gamma^M (x_M^a - x_M^b)\chi_T^{ab} + 
\overline{\chi}_T^{ab} \Gamma^M (x_M^a - x_M^b) \chi_S^{ab} \right\} \;\;\;.
\end{eqnarray}
where $a, b=1,\cdots,N$.
  Introducing
\begin{eqnarray}
\chi_U^{ab} \equiv \chi_S^{ab} -i \chi_T^{ab}\;\;, \;\;
\chi_L^{ab}  \equiv \chi_S^{ab} +i \chi_T^{ab}\;\;,\;\;
{\cal M}^{ab} = \Gamma^M (x_M^a - x_M^b) \;\;.
\end{eqnarray}
  we find
\begin{eqnarray}
{\cal S}
= \frac{1}{2}\sum_{a<b} \left\{ \overline{\chi}_U^{ab} {\cal M}^{ab} 
\chi_U^{ab} - 
\overline{\chi}_L^{ab} {\cal M}^{ab} \chi_L^{ab} \right\} \;\;.
 \label{Unfermion}
\end{eqnarray}
  This action  belongs to the type I action in the next subsection.

\subsection{the case of the $USp(2k)$ matrix model}
  
 Along the same line, the fermionic action in $USp(2k)$ matrix model can be
 decomposed. Here we pay most of  our attention to
 the closed string sector  of $USp(2k)$ model.
 The sector  which the fundamental representation belongs to
 has been discussed in \cite{CIK}, and will be 
reviewed briefly at the end of this subsection.
 The fermionic part of the action takes the same form as eq.~(3.1).
But now the matrix fermion $\Psi$ is decomposed into the adjoint and the
 antisymmetric representation as is described in  eqs.~(\ref{eq:psi}),
 (\ref{eq:adjanti}).  As for $X_{M}$,
\beq
X_M=diag(x^1_M,\cdots,x^k_M,\rho(x^1_M),\cdots,\rho(x^k_M))\;\;,
\eeq
where the $\rho$ is a projection:
\beq
\ba{ccc}
\rho:  x_\mu  &\rightarrow&  \;\;   -x_\mu \;\;, \;\; \mu=0,1,2,3,4,7 
\;\;\;,\\ 
  x_n   &\rightarrow&  \;\;   x_n \;\;, \;\;  n=5,6,8,9\;\;\;.
\ea
\eeq
The generators of {usp}(2k)  excluding the Cartan subalgebras are
\begin{eqnarray}
\label{eq:adjointbases} 
S_{ab} - S_{a+k,b+k} \;,\;&&    T_{ab} + T_{a+k,b+k} \;\;, \;\;
 a<b \in \{ 1,2,\cdots ,k \} \nonumber \\
S_{a,b+k} + S_{b,a+k}\; ,\; && T_{a,b+k} + T_{b,a+k} \;\;, \;\; 
a<b \in \{ 1,2,\cdots ,k \} \label{adjoint}\\
S_{a,a+k}\; ,\; && T_{a,a+k}\;\;, \;\;  a = 1,2, \cdots , k \;\;\nonumber
\end{eqnarray}
  while the generators of {asym}(2k)  excluding the diagonal ones are
\begin{eqnarray}
\label{eq:asymbases}  
S_{ab} + S_{a+k,b+k}  \;, \; &&  T_{ab} - T_{a+k,b+k}  \;\;, \;\;
 a<b \in \{ 1,2,\cdots ,k \} \nonumber \\
S_{a,b+k} - S_{b,a+k}  \; ,\; && T_{a,b+k} - T_{b,a+k} \;\;, \;\;
 a<b \in \{ 1,2,\cdots ,k \}\;\;.\label{asym}
\end{eqnarray}
 The fermions $\Psi_{adj}$ and $\Psi_{asym}$ are expanded respectively by
 the bases  eq.~(\ref{eq:adjointbases}) and  eq.~(\ref{eq:asymbases}).  
  
After some algebraic manipulation which we leave in the appendix,
   we find that the fermionic action of 
$USp(2k)$ reduced model is expressed in terms of three types of actions
\begin{eqnarray}
{\cal S}&=&\frac{1}{2}\{ \sum_{a<b}{\cal L}_{I}
\left(\left(\Psi_{adj}^{DU}\right)^{ab} ,
\left(\Psi_{asym}^{DU}\right)^{ab} ; x_M^a ,  x_M^{b}\right)  \nonumber \\
&+& \sum_{a<b}{\cal L}_{I}\left(\left(\Psi_{adj}^{DL}\right)^{ab} ,
 \left(\Psi_{asym}^{DL}\right)^{ab} ; -x_M^a  ,- x_M^{b}\right) \nonumber  \\
&+& \sum_{a<b}{\cal L}_{II}\left(\left(\Psi_{adj}^{OU}\right)^{ab} ,
 \left(\Psi_{asym}^{OU}\right)^{ab} ; x_M^a  , x_M^{b}\right)   \nonumber \\
&+& \sum_{a<b}{\cal L}_{II}\left(\left(\Psi_{adj}^{OL}\right)^{ab} ,
\left( \Psi_{asym}^{OL}\right)^{ab} ; - x_M^a ,  - x_M^{b}\right) \nonumber  \\
&+& \sum_{a}{\cal L}_{III}\left(\left(\Psi_{adj}^{ODU}\right)^{a} ;
 x_\mu^a - x_\mu^{a+k}\right) \nonumber  \\
&+& \sum_{a}{\cal L}_{III}\left(\left(\Psi_{adj}^{ODL}\right)^{a} ;
 -x_\mu^a + x_\mu^{a+k}\right)\}  \label{action} \;\;\;,
\end{eqnarray}
where
\begin{eqnarray}
{\cal L}_{I} \left( \Lambda, \Phi ; x_{M}, y_{M} \right)
 &\equiv& 2\left( \overline{\Lambda} + \overline{\Phi} \right) 
\Gamma^M ( x_M -y_M) \left( \Lambda + \Phi \right)  \;\;\;, \\ 
{\cal L}_{II} \left( \Lambda, \Phi ; x_{M}, y_{M} \right)
 &\equiv& 2\left( \overline{\Lambda} +
 \overline{\Phi} \right) \Gamma^M ( x_M -\rho(y_M)) 
\left( \Lambda + \Phi \right)  \;\;\;, \\
{\cal L}_{III}
\left( \Lambda ; x_{\mu} \right)
 &\equiv& \overline{\Lambda} \Gamma^\mu x_\mu \Lambda \;\;\;.
\end{eqnarray}
We call ${\cal L}_I, {\cal L}_{II}$ and ${\cal L}_{III}$ 
 type I, type II and type III action respectively.
  See the appendix for detail of our notation.

It is obvious that only components from the diagonal blocks $D$ contribute to
 type I action, while only components from $O$ contribute to type II action.
As for type III action,  only components from $OD$ contribute and they are
  all in the adjoint representation. We will 
find  that the contribution from the  fermions in the fundamental
 representation has the same form as type III action.
We see that the part of the adjoint fermions and all of
 the antisymmetric fermions form Majorana-Weyl fermions  while the remainder
of the adjoint fermions decouple from the spacetime points lying in the
 antisymmetric representation.
  We indicate below    the parts of the matrix degrees of freedom of the
 fermion  $\Psi$   constributing to  ${\cal L}_{I}$,
 ${\cal L}_{II}$ and ${\cal L}_{III}$ by
 $\bullet$, $\circ$,   and $\star$  respectively .
\begin{eqnarray}
\left(
\begin{array}{ccc|ccc}
  0&  \bullet&  \bullet&  \star&  \circ&  \circ\\
  \bullet&  0&  \bullet&  \circ&  \star&  \circ\\
  \bullet&  \bullet&  0&  \circ&  \circ&  \star\\\hline
  \star&  \circ&  \circ&  0&  \bullet&  \bullet\\
  \circ&  \star&  \circ&  \bullet&  0&  \bullet\\
  \circ&  \circ&  \star&  \bullet&  \bullet&  0
\end{array}
\right)
\end{eqnarray}
%\begin{figure}
%%\begin{figure}[t]
%\epsfysize=5cm 
%\vspace*{1cm}
%\centerline{\epsfbox{string5.eps}}
%\caption{}
%\label{total}
%\end{figure}

Apart from the fermions in the closed string sector, we also have
  the fermions belonging to the fundamental representation
in eqs.~(\ref{eq:gf}) and  (\ref{eq:Yukawa}).
The action,  up to the prefactor, reads
\beqn
{\cal S}_{F} = \psivector^{\ast}
  \overline{\sigma}^{m} X_{m} \cdot \psivector +
  \frac{1}{2}  \left(  \psivector  \cdot  \Sigma   F
  \left( X_{4}+ iX_{7} + M \right) \psivector
  +  h.c.   \right) \;\;. 
\eeqn
  As is already discussed in \cite{IM, CIK}, this action does not depend
 on $X_{n}$ $n=5,6,8,9$ and 
 is in fact type III;
\beqn
\label{actionF}
{\cal S}_{F} =  \sum_{a, f, \pm} {\cal L}_{III} \left( \psi_{Q_{(f)} a}
 ; x_{\mu}^{a} \pm m_{(f)} \delta_{\mu, 4} \right) \;\;.
\eeqn
See \cite{IM, CIK} for more detail. 
 The fermionic part of the action reads
\beqn
  S_{fermion} \sim   {\cal S} + {\cal S}_{F} \;\;.
\eeqn

\section{Computation of nonabelian Berry phase}
  We now proceed to the computation of the nonabelian Berry phase.
  Eqs. (\ref{eq:3a}) (\ref{eq:12a}) are our results.
\subsection{type III case}
The generic  type III action ${\cal L}_{III}$ in the last section
  do not really depend on the spacetime points $x_{n}$
 belonging to the antisymmetric representation. Let us first compute
 the nonabelian Berry phase for the Hamiltonian of this type.
  Putting back the prefactor discarded in the last section,
  we obtain the corresponding first quantized Hamiltonian:
\begin{equation}
{\cal H} = \frac{1}{g^2} \sum_{\mu=1,2,3,4,7} x_{\mu}  \gamma^{\mu}
 \;\;,
\end{equation}
  where $\gamma^{\mu}$s are the five dimensional gamma matrices obeying
 the Clifford algebra. We take the following representation:
\begin{eqnarray}  
\begin{array}{ccc}
\gamma^1 = \sigma^1 \otimes \sigma^3\;\;,&
\gamma^2 = \sigma^2 \otimes \sigma^3\;\;,&
\gamma^3 = \sigma^3 \otimes \sigma^3\;\;,\\
\gamma^4 = {\bf 1} ^2 \otimes \sigma^1\;\;,&
\gamma^7 = {\bf 1} ^2 \otimes \sigma^2  \;\;.
\end{array}
\end{eqnarray}
where $\s^i$ are the Pauli matrices.

 The eigenvalues of eq.~(4.1) are 
 ${\displaystyle \pm \frac{ \mid x \mid}{g^{2}} }$ with
 $\mid x \mid \equiv {\displaystyle \sqrt{ \sum_{\mu= 1,2,3,4,7} 
  x_{\mu}^{2} } }$  and each one is doubly degenerate. We focus on the
 two dimensional subspace of the one-particle states which belongs to the
 positive eigenvalue. The nonabelian Berry connection obtained will be
 $su(2)$ Lie algebra valued.
The eigenstates can be obtained with
the help of projection operators,  which are defined by
\begin{equation}
P_{\pm}  \equiv \frac{1}{2} ({\bf 1}_4 \pm  y_{\nu}
 \gamma^{\nu}) \;\;, \;\;
\end{equation}
  where
\beq 
 y_{\nu} \equiv \frac{x_{\nu}}{\mid x \mid} \;\;\;
\eeq
 parametrize ${\bf S}^4$.
 These projection operators satisfy
\beq
{P_{\pm}}^2 = P_{\pm} \hspace{3ex}
 P_+^{\dag} = P_+\hspace{4ex}{\cal H}P_{\pm}=\pm\frac{|x|}{g^2}P_\pm \;\;.
\eeq
 
Let us denote by ${\bf e}_\alpha$ $(\alpha = 1,2,3,4)$ the component
 representation of  the unit vector in the
 $i$-th direction, i.e., the one nonvanishing only at the i-th position
\beq
{\bf e}_\alpha  \equiv \trans{(\cdots,0,1,0,\cdots)} \;\;.
\eeq
The normalized eigenvectors with plus eigenvalue are
 \begin{equation}
 \psi_\alpha  = \frac{1}{{\cal N}_\alpha} P_{+}{\bf e}_\alpha \;\;\;.
\end{equation}
Here, $\psi_\alpha \;\;, \alpha= 1, 4 $ form a two-dimensional
 eigenspace well-defined around the north pole $x_{3}=
 \mid x \mid$ 
 while  $\psi_\alpha, \alpha= 2,3$ the one around the south pole
 $x_{3}=-\mid x \mid$ .  We see that the origin of the degeneracy is in the
 spinor index.
  The ${\cal N}_\alpha $ are the normalization factors :
\begin{equation}
{\cal N} \equiv {\cal N}_1 = {\cal N}_4 = \sqrt{\frac{1+y_3}{2}}
\;\;,\;\;
 {\cal N}^{\prime} \equiv{\cal N}_2 = {\cal N}_3 =
 \sqrt{\frac{1-y_3}{2}}\;\;.
\end{equation}

  We focus our attention on the sections near the north pole.
 The Berry connection \cite{Berry, WZ}  is
\begin{equation}
i{\cal A}  = \left( \begin{array}{l}
			  \psi_1 \\
			 \psi_4 
		\end{array} \right)
	 d (  \psi_1,  \:  \psi_4  )
 = {\cal E}{\cal M}\trans{\cal E} \;\;,
\end{equation}
 where 
\beqn
{\cal M} &\equiv&  \frac{1}{{\cal N} }P_+^{\dag} d 
 \frac{1}{{\cal N}}P_+  \;\;\;, \\  
{\cal E}  &=&   \trans{({\bf e}_1,{\bf e}_4)} \;\;\;.
\eeqn
  Introducing 
\beqn
C_{\mu \nu} \equiv (y_{\mu} dy_{\nu} -
 y_{\nu} dy_{\mu})\;\;, 
\eeqn
 we obtain
\begin{equation}
{\cal M}= \frac{1}{1 + y_3} \left( \frac{1}{2}dy_{\mu}
 \gamma^{\mu} +
\frac{1}{4}C_{\mu \nu} \gamma^{\mu} \gamma^{\nu} -
 \frac{1}{1 + y_3} dy_3 P_+ \right) \;\;,
\end{equation}
\begin{equation}
\label{NABCAL} 
{\cal A}  \left( y_i \right) = \frac{ 1 }{2(
 1+y_3)} {\bf B} \cdot
 \mbox{\boldmath $\sigma$} \;\;,
\end{equation}
where
\begin{equation}
{\bf B}   \equiv
 \left[
\begin{array}{l}
     B_1 \\
     B_2 \\
     B_3
\end{array}\right]   =
 \left[
\begin{array}{l}
y_7 dy_1 - y_1 dy_7 - y_2 dy_4 + y_4 dy_2 \\
y_1 dy_4 - y_4 dy_1 - y_2 dy_7 + y_7 dy_2 \\
y_4 dy_7 - y_7 dy_4 - y_2 dy_1 + y_1 dy_2
\end{array}\right]\;\;.
\end{equation}
  Observe that $y_3$ appears only in the overall scale factor.
Define
\begin{equation}
T \equiv \frac{1}{\sqrt{1-y^2_3}} \left( y_2 {\bf 1}_2 + i {\bf y} \cdot
 \mbox{\boldmath $\sigma$ } \right)\;\;.
\end{equation}
The Berry connection can then be rewritten as
\beq
\label{eq:3a}
{\cal A}(y_\nu)=\frac{1-y_3}{2}dT T^{-1} \;\;. \label{eq:BPST} 
\eeq

Eq.~(\ref{eq:BPST}) is  the form we have obtained  in \cite{CIK}.
 We now discuss our renewed understanding.  We will show  below that,
 by a further change of coordinates,
 the above non-Abelian connection $\cal A$  can be brought exactly to
  the form of the BPST instanton configuration \cite{BPST}. 
We are in the five dimensional space with coordinates
 $x_1, x_2, x_3, x_4, x_7$, and in this coordinate system
 the Berry connection seems to
 have a prefactor depending on $x_3$. As is seen in \cite{Yang}, however,
the connection is independent of the radius if we work in the polar coordinate
 system.   This means that we can consider the problem on ${\bf S}^4$.
From this point of view alone, $\cal A$ is a nontrivial $SU(2)$ bundle over
 ${\bf S}^4$ with second Chern number $\mp 1$. In other words, it should be
 the BPST self-dual or anti-selfdual(ASD) instanton connection with
 instanton number $\pm1$.
 To show this explicitly, we  make the following series of  change of
 coordinates: first change to  the polar coordinate system on ${\bf S}^4$ and,
 via the stereographic projection, change to
  the orthogonal coordinate system on ${\bf R}^4$ where the connection is made
 manifestly  the ASD $SU(2)$ connection.
The transformations  which realize these are found to be
\beq
y_i=\frac{2z_i}{1+z^2}  \;\;, \;\;\hspace{4ex}i=1,2,4,7 \;\;, \hspace{5ex}
y_3=\frac{1-z^2}{1+z^2} \;\;,
\eeq
where \{$z_i$\} \footnote{ In our notation $z_i$ are real numbers.}
 parameterize ${\bf R}^4$, and 
\beqn
T  &=&   \frac{1}{|z|}(z_2{\bf 1}_2+i {\bf z}\cdot \mbox{\boldmath$\sigma$})=
\frac{1}{|{\hat z}|}{\hat z}\;\; ,  \\
dTT^{-1} &=&   dT \bar{T}=\frac{1}{2|{\hat z}|^2}(d{\hat z}\bar{\hat  z}
-{\hat  z}d\bar{\hat  z}) \;\;.
\eeqn
Here ${\hat  z}$ is a quaternion
\beq
{\hat  z} \equiv z_2{\bf 1}_2+i {\bf z}\cdot \mbox{\boldmath$\sigma$} \;\;\;.
\eeq
We obtain
\bea
{\cal A} &=&\frac{1-y_3}{2}dT T^{-1}\nonumber\\
&=&\frac{z^2}{1+z^2}dT\bar{T}\nonumber\\
&=&\frac{1}{1+{\hat  z}^2}\cdot \frac{1}{2}(d{\hat  z}
\bar{\hat  z}-{\hat  z}d\bar{\hat  z}) \;\;,
\eea
which is exactly the gauge connection for the ASD instanton
 on  ${\bf R}^4$\cite{Atiyah}. 

In terms of {$z_i$}, the Berry connection can also be written as
\beq
\label{eq:asu(2)}
{\cal A}=\frac{1}{2}\cdot \frac{1}{1+z^2} (z_idz_j-z_jdz_i)\sigma^{ij} \;\;,
\eeq
where
\beq
\sigma^{ij}\equiv \frac{i}{2}{\cal E}[\gamma^i, \gamma^j]\trans{\cal E} \;\;,
\eeq
with components
\beq
\sigma^{12}=\left[ \ba{cc}
-1&0  \\
0&1
\ea \right]\; ,\hspace{4ex}\sigma^{14}=\left[ \ba{cc}
0&i\\
-i&0
\ea \right]  \;,
\eeq
\beq
\sigma^{17}=\left[ \ba{cc}
0&1\\
1&0
\ea \right]\;  ,\hspace{4ex}\sigma^{24}=\left[ \ba{cc}
0&1\\
1&0
\ea \right]   \;,\\
\eeq
\beq
\sigma^{47}=\left[ \ba{cc}
-1&0\\
0&1
\ea \right]  \;,\hspace{4ex}\sigma^{27}=\left[ \ba{cc}
0&-i\\
i&0
\ea \right]\; .
\eeq
The curvature two-form of $\cal A$ is 
\beq
{\cal F}=\frac{1}{(1+z^2)^2}dz_i\wedge dz_j\s^{ij} \;\;.
\eeq
As expected, the curvature satisfies the anti-selfdual condition
\beq
{\cal F}_{\mu\nu}+\frac{1}{2}\epsilon_{\mu\nu\s\rho}{\cal F}^{\s\rho}=0 \;\;,
\eeq 
and the instanton number, i.e., the second Chern class $k=-C_2(P)$ is
\beq
k=-\frac{1}{8\pi^2}Tr\int{\cal F}\wedge{\cal F}=-1 \;\;.
\eeq 
In the same way, the Berry connection corresponding to the minus
 eigenvalue can be evaluated. It is a selfdual instanton with $k=1$. 

For the sake of our discussion, it is more
 appropriate to regard this as pointlike nonabelian singularity
 located at the origin  of five space dimensions. In fact,   this is
 what is sometimes called Yang monopole \cite{Yang} or an
SU(2) monopole in five-dimensional flat space or
 four-dimensional spherical space. It has a nontrivial Chern number
 on ${\bf S}^{4}$ and its topological stability is summarized by
 $\pi_3(SU(2)) = {\bf Z}$  as is the case for the BPST instanton.

\subsection{type I, II}

 Let us turn to the evaluation of the nonabelian Berry phase associated with
 the action of type I, II.  The generic action and its Hamiltonian  are
respectively
\beqn 
{\cal L}_{I,II} &=&  - \frac{1}{g^{2}}
\bar{\Psi}\Gamma^M x_M\Psi \;\;, \;\;  {\rm and } \;\;,  \nonumber  \\
{\cal H}_{F} &=&  \frac{1}{g^{2}} \overline{\Psi} \Gamma^M x_M \Psi \;\;.
\eeqn
  Here $\Psi$  is  the ten-dimensional Majorana-Weyl spinor.
Working in the $v_0=0$ gauge,  the relevant first quantized Hamiltonian is
\begin{eqnarray}
{\cal H} =  \frac{1}{g^{2}} \sum_{i= 1,2,\cdots, 9} \Gamma^0 
\Gamma^i x_i \;\;\;.
\end{eqnarray}
As in the last subsection, we define a projection operator
\begin{eqnarray}
P_{\pm} \equiv \frac{1}{2}({\bf 1}_{16} \pm \frac{1}{|x|}
 \Gamma^0 \Gamma^i x_i) \;\;, \;\;
|x| \equiv  \sqrt{ \sum_{i=1}^9(x_i)^2  } \;\;.
\end{eqnarray}
This time, the projection operator acts on the Weyl projected
sixteen dimensional space. The gamma matrices are understood to act on
  this space and are regarded as $16 \times 16$ matrices.
With the help of this projection operator, the orthogonal eigenvectors can
 be constructed. 
In the case of the positive eigenvalue, the eigenvectors are
\begin{eqnarray}
  \psi_\alpha &=& \frac{1}{{\cal N}_\alpha} P_+ {\bf e}_\alpha \;\;,  \\
{\cal N}^2  &\equiv& {\cal N}_\alpha^2  = \frac{1}{2} \left( 1 +
 \frac{x_3}{|x|} \right)
\;\;, \;\; {\rm for} \;\; \alpha=1,3,5,7,10,12,14,16
 \;\;, \nonumber \\
{\cal N}^{\prime2} &\equiv& {\cal N}_\alpha^2 =  \frac{1}{2}
 \left( 1 - \frac{x_3}{|x|} \right)
\;\;, \;\; {\rm for} \;\;
 \alpha =2,4,6,8,9,11,13,15  \;\;.  \nonumber
\end{eqnarray}
  Here the second line  is well-defined around the north pole  while
  the third line is well-defined  around  the south pole. 

 We focus on the eigenspace well defined around the north pole  with the
  positive eigenvalue.  The eigenspace forms an eight dimensional vector
 space.  The nonabelian Berry phase is  $su(8)$ Lie algebra valued
 one form and is given by
\begin{eqnarray}
{\cal A} &=&-i
\left( \begin{array}{c}  \psi_1 \\  \psi_3 \\ \vdots \\  \psi_{14} 
 \\ \psi_{16} \end{array}
\right)
d \left( \psi_1, \; \psi_3, \; \cdots \psi_{14}\; \psi_{16}
 \right)  \;\;, \nonumber \\
&=&{\bf E}{\cal O}\trans{\bf E} \;\;,
\end{eqnarray}
where 
\begin{equation}
{\bf E} \equiv \trans{\left(  {\bf e}_{1},{\bf e}_{3},{\bf e}_{5},
{\bf e}_{7},{\bf e}_{10},{\bf e}_{12},{\bf e}_{14},{\bf e}_{16} \right)} \;\;,
\end{equation}
and 
\begin{eqnarray}
{\bf {\cal O} }& \equiv &-i \frac{1}{{\cal N}} P_+^{\dag}
 d \frac{1}{{\cal N}} P_+  \;\;,\\
 &=& -i \frac{1}{{\cal N}^2} \left( - \frac{1}{2 {\cal N}^2}
 d ({\cal N})^2  P_+ + P_+ d P_+ \right)  \;\;\nonumber.
\end{eqnarray}
  Introduce the coordinates
\begin{equation}
y_i   \equiv \frac{x_i}{|x|} \;\;
\end{equation}
  which parametrize ${\bf S}^{8}$.
 We find
\begin{eqnarray}
\label{eq:Oberry}
{\bf {\cal O} }
&=& -i \frac{1}{1+y_3} \left( - \frac{1}{(1+y_3)} 
  P_+ d y_3+ P_+  \Gamma^0 \Gamma^i d y_i \right) \;\;,  \nonumber  \\
&=& -i \frac{1}{2(1+y_3)} \left\{  \frac{1}{(1+y_3)}\left( - 
  {\bf 1}_{16}   d y_3 -  \Gamma^0 \Gamma^i y_i d y_3+  
 (1+y_3)\Gamma^0 \Gamma^i d y_i \right) \right.  \;\; \nonumber \\
  & &   \left. - 
 \frac{1}{4}\left[  \Gamma^i , \Gamma^j \right] C_{ij}
 \right\} \;\;.
\end{eqnarray}
Here  we have introduced
\begin{equation}
C_{ij} \equiv y_idy_j -y_jdy_i \;\;,
\end{equation}
 and  used
\beq
 \sum_{i= 1,2,\cdots, 9} y_i dy_i =0 \;\;.
\eeq
 Observe  that
\begin{eqnarray}
&& -   {\bf 1}_{16}   d y_3 -  \Gamma^0 \Gamma^i y_i d y_3+ 
  (1+y_3)\Gamma^0 \Gamma^i d y_i   \nonumber   \\
&=&  ({\bf 1}_{16}+\Gamma^0\Gamma^3)dy_3+\sum_{A=1,2,4,5,6,7,8,9}
 \Gamma^0\Gamma^A \left(  
 dy_A +  C_{3A} \right) \;\;,
\end{eqnarray}
   as well as
\beq
{\bf E}\left\{ ({\bf 1}_{16}+\Gamma^0\Gamma^3)dy_3+ \sum_{A}
 \Gamma^0\Gamma^A \left(   dy_A +  C_{3A} \right) \right\} \trans{\bf E}=0
\;\;,
\eeq
due to the orthogonality of different eigenvectors.  By the same reason,
  the last line in eq.~(\ref{eq:Oberry}) with  $y_3$ term involved will not
 contribute to the Berry connection either.  We conclude that
 only the last term of ${\cal O}$  with $i,j \neq 3$
  contributes to the nonabelian Berry phase.  

Notice that the vector $\bf E$ satisfies
\begin{eqnarray}
\trans{\bf E}{\bf E} &=& \frac{1}{2}\left( 1+ \Gamma^0\Gamma^3 \right)
 \;\;, \\
\Gamma^0\Gamma^i \trans{\bf E}{\bf E} + \trans{\bf E}{\bf E}
 \Gamma^0\Gamma^i &=& 
\Gamma^0\Gamma^i\;\;, \;\;\; {\rm for} \;\;  i=1,2,4,5,6,7,8,9 \;\;.
\end{eqnarray}
  Defining $8\times 8$ matrix
\begin{equation}
\Sigma^{ij} \equiv \frac{i}{2}{\bf E} [ \Gamma^i , \Gamma^j ] \trans{\bf E}
\;\;,
\end{equation}
  we obtain
\begin{equation}
\label{eq:12a}
{\cal A} =  \frac{1}{4(1+y_3)}  C_{ij} \Sigma^{ij}\;\;.
\end{equation}
  Here the indices $i,j$ run over the eight directions.
 The generators  $\Sigma^{ij}$  is found to  form
 an $so(8)$ algebra:
\begin{eqnarray}
[ \Sigma^{ij} , \Sigma^{kl} ] &=& 2i \left( \delta^{jk}\Sigma^{il} 
- \delta^{jl}\Sigma^{ik} - 
\delta^{ik}\Sigma^{jl} + \delta^{il}\Sigma^{jk} \right) \;\;.
\end{eqnarray}
The curvature  is 
\begin{eqnarray}
{\cal F}& = &d {\cal A} - i {\cal A}\wedge 
{\cal A}  \;\;, \nonumber \\
&=&\frac{1}{4}dy_i \wedge dy_j \Sigma^{ij}
 - \frac{1}{4(1+y_3)} dy_3 \wedge C_{ij} \Sigma^{ij}  \;\;.
\end{eqnarray}

As in the case of the $SU(2)$ Berry connection, we change our coordinate
 system  to that on ${\bf R}^8$:
\beq
y_i=\frac{2z_i}{1+z^2}, \hspace{5ex}y_3=\frac{1-z^2}{1+z^2} \;\;.
\eeq
  Here $z_i \;,\; (i=1,2,4,\cdots,9)$  are the orthogonal coordinates on
 ${\bf R}^8$. The Berry connection can be rewritten as
\beq
\label{eq:asu(8)}
{\cal A}  =\frac{1}{ 2\left(1+z^2 \right)}
(z_idz_j-z_jdz_i)\Sigma^{ij}  \;\;.
\eeq
The curvature of this connection  is
\beq
\label{eq:F8}
 {\cal F} = \frac{1}{(1+z^2)^2}dz_i\wedge dz_j\Sigma^{ij} \;\;.
\eeq
  Let us see if this eigenbundle is nontrivial.
 The nontrivial $SU(8)$ bundle on ${\bf S}^8$  is characterized by the seventh
 homotopy group
\beq
\Pi_7(SU(8))= {\bf Z}\;\;.
\eeq
This leads us to compute the fourth Chern number  using
  eq.~(\ref{eq:F8}) derived from our eigenbundle:
\beq
C_4(P)=\int c_4(P)=(\frac{i}{2\pi})^4\cdot\frac{1}{4!}\int
 Tr({\cal F}^4)=1 \;\;.
\eeq
 Our eigenbundle  is in fact nontrivial and has the fourth Chern number $1$.

\section{Spacetime picture emerging from our computation }
  Let us recall the generic formula (eq.~(\ref{eq:formulaPB}) ) stated in the
 end of section two:
\beqn
     \langle \langle \;\;
 \hat{{\cal P}}_{\ell}^{\alpha \alpha^{\prime}}\;\; \rangle \rangle_{\Gamma}
  =
 P \exp \left[  -i  \oint_{\Gamma} {\cal A}_{\ell} \left( x_{M} \right)
    \right]^{\alpha \alpha^{\prime}} \;\;.
\eeqn
The results from our computation in the last section  are  summarized as
 eqs.~(\ref{eq:3a}), (\ref{eq:asu(2)}):
\beqn
{\cal A}_{SU(2)}   &=&  \frac{1-y_3}{2}dT T^{-1} \;\;,  \nonumber \\
{\cal A}_{SU(2)}  &=&  \frac{1}{2}\cdot \frac{1}{1+z^2} 
(z_idz_j-z_jdz_i)\sigma^{ij}
 \;\;\;, \nonumber   \\
\sigma^{ij} &\equiv&  \frac{i}{2}{\cal E}[\gamma^i, \gamma^j]\trans{\cal E}
\;\;\; , \nonumber
\eeqn
 for the generic type III action ${\cal L}_{III}$, giving
 the $SU(2)$ monopole

\noindent
 and as eqs.~(\ref{eq:12a}), (\ref{eq:asu(8)}):
\beqn
{\cal A}_{SU(8)} &=&  \frac{1}{4(1+y_3)}  C_{ij} \Sigma^{ij}\;\;, \nonumber \\
{\cal A}_{SU(8)}  &=& \frac{1}{2}\cdot\frac{1}{1+z^2}(z_idz_j-z_jdz_i)
\Sigma^{ij}  \nonumber
\;\;,\\
\Sigma^{ij} &\equiv& \frac{i}{2}{\bf E} [ \Gamma^i , \Gamma^j ] \trans{\bf E}
\;\;,  \nonumber
\eeqn
 for the generic type I,II action ${\cal L}_{I,II}$, giving
  the $SU(8)$ monopole. Putting these together, we state that
 the expectation value of the projector of a fermionic eigenmode is given by
  the path-ordered exponential of the integration of the connnection one-form
 and that this factor   in the case of $USp$ matrix model is controlled by
 the $SU(2)$ or the $SU(8)$ nonabelian monopole singularity sitting at the
 origin of the parameter space $X_{M}$.
 In the case of the $SU(8)$ monopole, it is a pointlike singularity in nine
 dimensions while in the case of the $SU(2)$ Yang monopole it is a singularity
 which does not depend on the four antisymmetric directions.  The latter one,
 viewed as a singularity in the
 entire space, is not pointlike but is actually a four dimensionally
 extended object.
  The emergence of these interesting objects, albeit being aposteriori,
 justifies the study of this expectation value  rather than of the fermionic
 part of the partition function.

  The matrix models in general contain many species of fermions, which
  couple to different spacetime points or $D0$ branes.
  They provide a collection of nonabelian Berry phases rather than just
 one.   With this respect, it is more appropriate to consider the manybody
   counterpart indicated in section two  (eq.~(\ref{eq:prodob}) ):
\beqn
\label{eq:prodanswer}
   \langle \langle \;\;
  \prod_{\ell \in {\cal I}_{+} }
  \sum_{\alpha} \hat{{\cal P}}_{\ell}^{\alpha \alpha}\;\;
 \rangle \rangle_{\Gamma}   =
   \langle \langle \;\;  \prod_{\ell \in {\cal I}_{+} }
  \sum_{\alpha} \hat{{\cal P}}_{\ell}^{\alpha \alpha}\;\;
 \rangle \rangle_{\Gamma} =  \prod_{\ell \in {\cal I}_{+} } 
 tr_{\ell} P \exp \left[  -i  \oint_{\Gamma} {\cal A}_{\ell} 
\left( x_{M} \right) \right] \;\;,
\eeqn
 where the subset ${\cal I}_{+}$ of all eigenmodes is taken over the eigenmodes
 belonging to the positive eigenvalues.
 
  Actually, $S_{fermion} \sim {\cal S} + {\cal S}_{F}$ of the $USp$ matrix
 model contains many terms consisting of the generic ${\cal L}_{I,II}$ type
 action as well as the generic ${\cal L}_{III}$ type action.
   Listing the parameters which they depend on, we obtain from
  eq.~(\ref{action})
\beqn
\label{eq:list11}
{\rm  i)}&\;\;  x_{M}^{a} -x_{M}^{b} \;\;, 
 - \left( x_{M}^{a} -x_{M}^{b} \right) \;\; \nonumber \\
 {\rm  ii)} &\;\; x_{M}^{a} -\rho \left( x_{M}^{b} \right) \;\;,
 - \left( x_{M}^{a} -  \rho \left( x_{M}^{b} \right)
 \right) \;\;, \nonumber \\
 {\rm iii)} &\;\;  2x_{\mu}^{a} \;\;,    -2x_{\mu}^{a}  \;\;.
\eeqn
 Singularities occur when  the two points $x_{M}^{a}$ and
 $x_{M}^{b}$ collide in the case of the first line,
  when  $x_{M}^{a}$ and $\rho \left( x_{M}^{b} \right)$
 collide in the case of the second line and
 when  $x_{M}^{a}$ lies in the orientifold surface.
  Similarly from eq.~(\ref{actionF}),  we obtain 
\beqn
\label{eq:list22}
  {\rm  iv)} \;\;  x_{\mu}^{a} \pm m_{(f)} \delta_{\mu 0}
\eeqn
  The singularities occur when  $x_{\mu}^{a}$ is away from the
 orientifold surface in the $x_4$ direction by $ \pm m_{(f)}$.
  The situation is depicted in Figure 1.
\begin{figure}
%%\begin{figure}[t]
\epsfysize=5cm 
\vspace*{1cm}
\centerline{\epsfbox{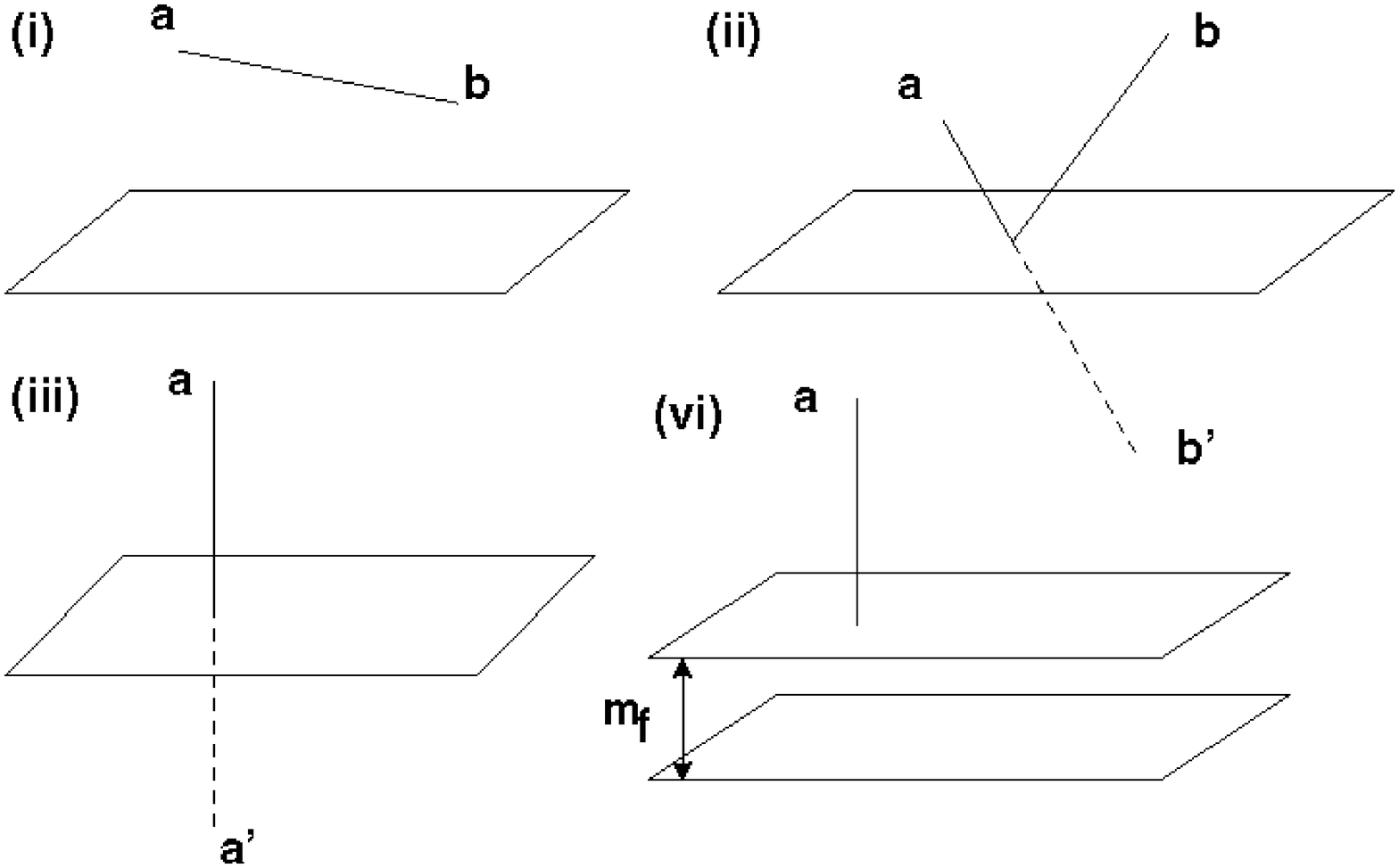}}
\caption{}
\label{total}
\end{figure}

  We denote the first six
cases of eq.~(\ref{eq:list11}) by $x_{M}^{a,b (K)} \;\; K=1, \cdots 6 $.
  and the last two cases of eq.~(\ref{eq:list11}) and
  the case of eq.~(\ref{eq:list22}) ( sum over
  $a$ and $f$ ) by  $x_{\mu}^{a (K^{\prime})}$.
  In the case where more than two spacetime points collide, we will
 obtain an enhanced symmetry which supports a nonabelian monopole.

 As for eq.(\ref{eq:prodanswer}), the subset  ${\cal I}_{+}$ is taken
 over the eigenmodes seen in eq.~(\ref{eq:list11}) and eq.~(\ref{eq:list22}).
 To write this more explicitly,
\beqn
 \prod_{K} \prod_{a < b} tr_{8}
 P \exp \left[  -i  \oint_{\Gamma} {\cal A}_{SU(8)}
 \left( x_{M}^{a,b (K)} \right) \right]
 \prod_{K^{\prime}} \prod_{a} tr_{2}
 P \exp \left[  -i  \oint_{\Gamma} {\cal A}_{SU(2)}
 \left( x_{\mu}^{a (K^{\prime})} \right) \right] \;\;.
\eeqn

 In the context of a matrix model for unified theory of superstrings, 
 measuring  eq.~(\ref{eq:prodob}), or eq.~(\ref{eq:formulaPB}) will provide
 means to examine spacetime formation suggested by the model.
 The presence of colliding singularities of spacetime points in general means
a dominant probability to such configurations.
 As we have said, there are two varieties of singularities
 we have exhibited in this paper. 
  The singularities of the $SU(8)$ monopoles appear
 to be evenly distributed in all directions as long as the off-diagonal bosonic
 integrations are ignored. This type is present both in the $IIB$ matrix
 model and the $USp$ matrix model.
The singularities of the $SU(2)$ monopoles  appear to be a string soliton
 of four dimensional extension  present in the USp matrix model and is not
 shared by the IIB matrix model.  It is tempting to think that
  the four dimensional structure  is formed by  a collection of  
 the $SU(2)$ monopoles.
 We put the spacetime picture emerging from contributions
 of various eigenmodes of the $USp$ matrix model as a table.  
\begin{eqnarray}
\begin{array}{|c|ccccccccc|}
\hline
&1&2&3&4&5&6&7&8&9\\
\hline
{\rm adjoint (D) + antisymmetirc (D)}&\times&\times&\times&\times&\times&
\times&\times&\times&\times\\
\hline
{\rm adjoint (O)+ antisymmetric (O)} &\times&\times&\times&\times&
\times&\times&\times&\times&\times\\
\hline
{\rm adjoint (OD)} &\times&\times&\times&\times&
\bigcirc&\bigcirc&\times&\bigcirc&\bigcirc\\
\hline
{\rm fundamental } &\times&\times&\times&\times&\bigcirc&
\bigcirc&\times&\bigcirc&\bigcirc\\
\hline\hline
{\rm orientifold }  &\times&\times&\times&\times&
\bigcirc&\bigcirc&\times&\bigcirc&\bigcirc\\
\hline
\end{array}
\end{eqnarray}

  As for the  $v_{0}$ direction, we have used this to parametrize the path
  in the remaining nine directions. The price we have to pay is that
 we have nothing to say on the  distributions of the spacetime points
  in this direction and this is closely related to the problem of the scaling
 limit.  Further progress and understanding of the spacetime formation
  of matrix models require overcoming this point.

\section{Generalization of the Yang monopole to odd dimensions}
  
In section IV, we have discussed the two kinds of nonabelian Berry
 connections  which have led us to the nontrivial $SU(2)$ and $SU(8)$ vector
 bundles over the parameter space. 
 The tools which have brought us to these bundles are the first quantized
 Hamiltonian with gamma matrices and the orthgonal projection operators.
 Letting  aside the physics context, these are certainly generalizable to
 arbitrary odd dimensions.
  In fact, the relevant mathematical discussion of such eigenbundle
  can be found for instance in \cite{Gilkey}. 

Let $\Gamma: {\bf R}^{m+1}
\rightarrow End( C^k )$ be a linear map. We assume 
\beq
\Gamma(x)^2=|x|^2\cdot I_k\hspace{2ex} {\rm and} \hspace{3ex} 
\Gamma(x)^{\ast}=\Gamma(x) \;\;.
\eeq
Let
\beq
\Gamma(x)=x_i\Gamma^i  \;\;.
\eeq
 Such {$\Gamma^i$}  satisfy the Clifford commutation rules:
\beq
\Gamma^i\Gamma^j+\Gamma^j\Gamma^i=2\delta^{ij} \;\;.
\eeq
That is,  $\Gamma^i$ are in fact gamma matrices. It is said that $\Gamma^{i}$
 give a ${\rm Cliff}(R^{m+1})$ module structure to $C^k$. 

Let 
\beq
P_{\pm}=\frac{1}{2}(1\pm \Gamma(x))\;\;,\hspace{3ex}  {\rm for}
 \hspace{2ex}|x|=1
\;\;.
\eeq
be an orthogonal projection onto the $\pm 1$ eigenspace of $\Gamma(x)$.
 Let
\beq
\Pi^{\Gamma}_{\pm}=\{(x,\nu)\in S^m\times C^k: \Gamma(x)\nu=\pm \nu\}
\eeq
be the corresponding eigenbundle. Clearly 
\beq
 S^m\times C^k=\Pi^{\Gamma}_{+}\bigoplus
\Pi^{\Gamma}_{-} \;\;.
\eeq
It is obvious that the orthogonal projection defined here is exactly the
 projection operator 
 defined from our first quantized Hamiltonian. 
 The Berry connection is just the connection
on  eigenbundle $\Pi^{\Gamma}_{+}$ or $\Pi^{\Gamma}_{-}$.
 It has been shown that the 
Chern number of this eigenbundle is 
\beq
\int_{S^{2j}}c_j(\Pi^{\Gamma}_{+})=i^j2^{-j}Tr(\Gamma^0\cdots\Gamma^{2j}) \;\;.
\eeq
 From this formula on the Chern number, we see that when $m$  is 
odd, $Tr(\Gamma^0\cdots\Gamma^m)$ vanishes and the eigenbundle is trivial.
When $m$ is even,    we find
\beq
Tr(\Gamma^0\cdots\Gamma^m)=2^{\frac{m}{2}}(-i)^{\frac{m}{2}}\;\;.
\eeq
Therefore the $\frac{m}{2}$-th Chern number should be 1.
  We have proven this explcitly by deriving
 the $SU(2)$ Berry connection on ${\bf R}^5$ ( the  $m=4$ case), and  the
 $SU(8)$ Berry connection on ${\bf R}^9$ (the $m=8$ case),
 and the curvature two-forms associated with them.
 The $U(1)$ Berry connection on ${\bf R}^3$  (the $m=2 $ case) is well known.
 The Berry connection on ${\bf R}^7$ (the $m=6$ case) may be related to
 some configuration in the reduced matrix model.
 
  Having formulated the eigenbundles associated with  the nonabelian
  Berry phase in arbitrary odd dimensions, we can safely state that
 the connnection one form and the curvatuire two form on ${\bf R}^{m+1}$
  are given by
\beq
\label{eq:aodd}
{\cal A}  =\frac{1}{ 2\left(1+z^2 \right)}
(z_idz_j-z_jdz_i)\Sigma^{ij}  \;\;.
\eeq
  and
\beq
\label{eq:Fodd}
 {\cal F} = \frac{1}{(1+z^2)^2}dz_i\wedge dz_j\Sigma^{ij} \;\;
\eeq
 respectively.
  Here $z_{i}$ are orthogonal coordinates on ${\bf R}^{m}$.

\section{Acknowledgements}
  We thank  Toshihiro Matsuo and Asato Tsuchiya  for helpful discussions,
  Soo-Jong Rey and Arkady Tseytlin  for interesting remarks,
 and Ashoke Sen and Ke Wu for useful comments.

\newpage

\appendix
\section{}
%\input{antsym3.tex}
%J
  In this appendix, we give some details of the derivation of the component
 expression of the action (eq.~(\ref{action}).  Let us first expand
 $\Psi_{adj}$ and $\Psi_{asym}$   by the generators stated respectively in 
 eq.~(\ref{adjoint}) and in eq.~(\ref{asym}):
\begin{eqnarray}
\Psi_{adj} &=& \sum_{a=1}^{k} \left\{ \left( \Psi_{adj}^{ODS}
 \right)^{a}S_{a,a+k} + 
 \left( \Psi_{adj}^{ODT} \right)^{a} T_{a,a+k} \right\}  \nonumber \\
&+&\sum_{a<b} \left\{ \left( \Psi_{adj}^{DS} \right)^{ab}\left( S_{ab} 
- S_{a+k,b+k} \right)  + \left( \Psi_{adj}^{DT} \right)^{ab}
 \left(T_{ab} + T_{a+k,b+k} \right) \right.   \nonumber \\
 &+& \left. \left( \Psi_{adj}^{OS} \right)^{ab}
 \left( S_{a,b+k} + S_{b,a+k} \right) + \left( \Psi_{adj}^{OT} \right)^{ab}
 \left(  T_{a,b+k} + T_{b,a+k} \right)\right\} \;\;,  \label{eq:ex1}  \\
\Psi_{asym} &=& \sum_{a<b} \left\{ \left( \Psi_{asym}^{DS}
 \right)^{ab}\left( S_{ab} + S_{a+k,b+k} \right) 
 + \left( \Psi_{asym}^{DT} \right)^{ab} \left(T_{ab} - T_{a+k,b+k} \right)
 \right.  \nonumber \\
 &+& \left. \left( \Psi_{asym}^{OS} \right)^{ab} 
\left( S_{a,b+k} - S_{b,a+k} \right) + 
\left( \Psi_{asym}^{OT} \right)^{ab} \left(  T_{a,b+k} - T_{b,a+k}
 \right)\right\} \;\;.  \label{eq:ex2} 
\end{eqnarray}
 Let us explain our notation  more carefully.
 $\Psi_{adj}$ is expanded by eq.~(\ref{adjoint}) which consists of three
sets of generators: the first set of generators ( the first line of
 eq.~(\ref{adjoint}) )  is  in the off-diagonal elements of
 the diagonal blocks ; the second set of generators ( the second line of
 eq.~(\ref{adjoint}) ) is in the off-diagonal elements of
 off-diagonal blocks;  and the third set of generators (the third line in
 eq.~(\ref{adjoint}))  is in the diagonal elements
 of the off-diagonal blocks.  We, therefore, distinguish these generators by
 $D$( the diagonal blocks), $O$( off-diagonal elements of the off-diagonal
 blocks) and $OD$( diagonal elements in the off-diagonal blocks) respectively.
 In each set of  the generators, there are contributions from both
    the real part $S$ and the purely
 imaginary part  $T$. We distinguish these by the superscript $S$ and $T$.
 
 As for $\Psi_{asym}$, the bases  are  $O$ type and $D$ type only. 
 In the above expansion, ( eqs.~(\ref{eq:ex1}), (\ref{eq:ex2})  ),
 the component
  fields are specified by the superscript specifying the species of
 the generators and the subscript  specifying the representation.
 For example, $\left( \Psi^{ODS}_{adjoint} \right)$ are the component fields
 of the adjoint fermions which correspond to the expansion coefficients of
  the $S$ type $OD$ generators. The same is true for the other components. 
%\begin{eqnarray*}
%&&\mbox{DS means real component of diagonal block.}\\
%&&\mbox{DT means imaginary component of diagonal block.}\\
%&&\mbox{OS means real component of off-diagonal block.}\\
%&&\mbox{OT means imaginary component of off-diagonal block.}\\
%&&\mbox{ODS means real diagonal component of off-diagonal block.}\\
%&&\mbox{ODS means imaginary diagonal component of off-diagonal block.}
%\end{eqnarray*}

 Let us denote 
\beq
-i(\bar{\psi}^S{\cal M}\psi^T-\bar{\psi}^T{\cal M}\psi^S)=
2Im(\bar{\psi}^S{\cal M}\psi_T)  \;\;.
\eeq
The component action can then be written as
\begin{eqnarray}
{\cal S}&=& 4\mbox{Im}\sum_{a<b}\left\{ \left( \overline{\Psi}_{adj}^{DS}
 \right)^{ab} 
+ \left( \overline{\Psi}_{asym}^{DS} \right)^{ab} \right\}{\cal M}^{ab}
 \left\{  \left( \Psi_{adj}^{DT} \right)^{ab} + \left( \Psi_{asym}^{DT}
 \right)^{ab} \right\}   \nonumber     \\
&+& 4\mbox{Im}\sum_{a<b}\left\{ -\left( \overline{\Psi}_{adj}^{DS}
 \right)^{ab} + 
\left( \overline{\Psi}_{asym}^{DS} \right)^{ab} \right\}{\cal M}^{a+k,b+k}
 \left\{  \left( \Psi_{adj}^{DT} \right)^{ab} - \left( \Psi_{asym}^{DT}
 \right)^{ab} \right\} \nonumber     \\
&+& 4\mbox{Im}\sum_{a} \left( \overline{\Psi}_{adj}^{ODS} \right)^{a} 
 {\cal M}^{a,a+k}
  \left( \Psi_{adj}^{ODT} \right)^{a}   \nonumber \\
&+& 4 \mbox{Im}\sum_{a<b}\left\{ \left( \overline{\Psi}_{adj}^{OS}
 \right)^{ab} +
 \left( \overline{\Psi}_{asym}^{OS} \right)^{ab} \right\}{\cal M}^{a,b+k}
 \left\{  \left( \Psi_{adj}^{OT} \right)^{ab} +
 \left( \Psi_{asym}^{OT} \right)^{ab} \right\} \nonumber   \\
&+& 4 \mbox{Im}\sum_{a<b}\left\{ \left( \overline{\Psi}_{adj}^{OS}
 \right)^{ab} 
- \left( \overline{\Psi}_{asym}^{OS} \right)^{ab} \right\}{\cal M}^{b,a+k}
 \left\{  \left( \Psi_{adj}^{OT} \right)^{ab} -
 \left( \Psi_{asym}^{OT} \right)^{ab} \right\} \;\;.
\end{eqnarray}
Here ${\cal M}^{ab}=x^a - x^b$.
Let us redefine the component fields, introducing complex notation. 
\begin{eqnarray}
\Psi_{adj}^{DU} &\equiv& \Psi_{adj}^{DS} - i \Psi_{adj}^{DT} \;\;, \;\;
\Psi_{adj}^{DL} \equiv \Psi_{adj}^{DS} + i \Psi_{adj}^{DT} \;\;, \;\;
\Psi_{asym}^{DU} \equiv \Psi_{asym}^{DS} - i \Psi_{asym}^{DT} 
 \;\;,  \nonumber \\
\Psi_{asym}^{DL} &\equiv& \Psi_{asym}^{DS} + i \Psi_{asym}^{DT}\;\;, \;\;
\Psi_{adj}^{ODU} \equiv \Psi_{adj}^{ODS} - i \Psi_{adj}^{ODT}\;\;, \;\;
\Psi_{adj}^{ODL} \equiv \Psi_{adj}^{ODS} + i \Psi_{adj}^{ODT}
  \;\;, \nonumber \\
\Psi_{adj}^{OU} &\equiv& \Psi_{adj}^{OS} - i \Psi_{adj}^{OT}\;\;, \;\;
\Psi_{adj}^{OL} \equiv \Psi_{adj}^{OS} + i \Psi_{adj}^{OT}\;\;, \;\;
\Psi_{asym}^{OU} \equiv \Psi_{asym}^{OS} - i \Psi_{asym}^{OT}\;\;, 
 \nonumber  \\
\Psi_{asym}^{OL} &\equiv& \Psi_{asym}^{OS} + i \Psi_{asym}^{OT}\;\;.
\end{eqnarray}
  We obtain
\begin{eqnarray}
{\cal S}&=& \sum_{a<b}\left\{ \left( \overline{\Psi}_{adj}^{DU} \right)^{ab}
 + \left( \overline{\Psi}_{asym}^{DU} \right)^{ab} \right\}{\cal M}^{ab}
\left\{  \left( \Psi_{adj}^{DU} \right)^{ab} +
 \left( \Psi_{asym}^{DU} \right)^{ab} \right\} \nonumber   \\
&+& \sum_{a<b}\left\{ -\left( \overline{\Psi}_{adj}^{DU} \right)^{ab}
 + \left( \overline{\Psi}_{asym}^{DU} \right)^{ab} \right\}{\cal M}^{a+k,b+k}
 \left\{  \left( \Psi_{adj}^{DU} \right)^{ab}
 - \left( \Psi_{asym}^{DU} \right)^{ab} \right\} \nonumber   \\
&-& \sum_{a<b}\left\{ \left( \overline{\Psi}_{adj}^{DL} \right)^{ab}
 + \left( \overline{\Psi}_{asym}^{DL} \right)^{ab} \right\}{\cal M}^{ab}
 \left\{  \left( \Psi_{adj}^{DL} \right)^{ab} +
 \left( \Psi_{asym}^{DL} \right)^{ab}
 \right\} \nonumber   \\
&-&\sum_{a<b}\left\{ -\left( \overline{\Psi}_{adj}^{DL} \right)^{ab}
 + \left( \overline{\Psi}_{asym}^{DL} \right)^{ab} \right\}{\cal M}^{a+k,b+k}
 \left\{  \left( \Psi_{adj}^{DL} \right)^{ab}
 - \left( \Psi_{asym}^{DL} \right)^{ab} \right\} \nonumber    \\
&+& \sum_{a} \left( \overline{\Psi}_{adj}^{ODU} \right)^{a} 
 {\cal M}^{a,a+k}  \left( \Psi_{adj}^{ODU} \right)^{a}   \nonumber \\
&-&\sum_{a} \left( \overline{\Psi}_{adj}^{ODL} \right)^{a} 
 {\cal M}^{a,a+k}  \left( \Psi_{adj}^{ODL} \right)^{a} \nonumber  \\
&+&\sum_{a<b}\left\{ \left( \overline{\Psi}_{adj}^{OU} \right)^{ab} 
+ \left( \overline{\Psi}_{asym}^{OU} \right)^{ab} \right\}{\cal M}^{a,b+k}
 \left\{  \left( \Psi_{adj}^{OU} \right)^{ab}
 + \left( \Psi_{asym}^{OU} \right)^{ab} \right\} \nonumber   \\
&+&\sum_{a<b}\left\{ \left( \overline{\Psi}_{adj}^{OU} \right)^{ab}
 - \left( \overline{\Psi}_{asym}^{OU} \right)^{ab} \right\}{\cal M}^{b,a+k}
 \left\{  \left( \Psi_{adj}^{OU} \right)^{ab}
 - \left( \Psi_{asym}^{OU} \right)^{ab} \right\} \nonumber  \\
&-& \sum_{a<b}\left\{ \left( \overline{\Psi}_{adj}^{OL} \right)^{ab}
 + \left( \overline{\Psi}_{asym}^{OL} \right)^{ab} \right\}{\cal M}^{a,b+k}
 \left\{  \left( \Psi_{adj}^{OL} \right)^{ab}
 + \left( \Psi_{asym}^{OL} \right)^{ab} \right\} \nonumber  \\
&-&\sum_{a<b}\left\{ \left( \overline{\Psi}_{adj}^{OL} \right)^{ab} - 
\left( \overline{\Psi}_{asym}^{OL} \right)^{ab}
 \right\}{\cal M}^{b,a+k} \left\{
  \left( \Psi_{adj}^{OL} \right)^{ab} - 
\left( \Psi_{asym}^{OL} \right)^{ab} \right\}\;\;.
\end{eqnarray}

Introduce three types of actions consisting of fermion bilinears:
\begin{eqnarray}
{\cal L}_{I}(\Lambda , \Phi ; x_M , y_M) &\equiv&\left( \overline{\Lambda} 
+ \overline{\Phi} \right) \Gamma^M (x_M - y_M) \left( \Lambda + \Phi \right)
 \nonumber \\
& & -\left( \overline{\Lambda} - \overline{\Phi} \right) \Gamma^M 
(\rho(x_M)-\rho(y_M)) \left( \Lambda - \Phi \right) \;\;,   \\
{\cal L}_{II} (\Lambda , \Phi ; x_M , y_M)&\equiv& 
 \left( \overline{\Lambda}  + \overline{\Phi}\right)
\Gamma^M(x_M-\rho(y_M))\left( \Lambda+\Phi)\right)
\nonumber \\
& &\mbox{}- \left( \overline{\Lambda} - 
\overline{\Phi} \right) \Gamma^M (\rho(x_M)- y_M) \left( \Lambda - 
\Phi \right)\;\;,    \\
{\cal L}_{III}(\Lambda ; x_M) &\equiv& 
\overline{\Lambda} \Gamma^{\nu} x_{\nu} \Lambda\;\;.
\end{eqnarray}

 Notice the relation
\begin{eqnarray}
\Gamma^0 \Gamma^M( x_M -y_M)  - \Gamma^0 H \Gamma^M (\rho(x_M)- 
\rho(y_M)) H =  2 \Gamma^0 \Gamma^M ( x_M -y_M)  \;\;,
\end{eqnarray}
with
\begin{eqnarray}
H \equiv \mbox{diag}(1,1,1,1,-1,-1,-1,-1,1,1,1,1,-1,-1,-1,-1) \;\;.
\end{eqnarray}
The above three types are simplified to be  
\begin{eqnarray}
{\cal L}_{I} &\equiv& 2\left( \overline{\Lambda} + \overline{\Phi} \right) 
\Gamma^M ( x_M -y_M) \left( \Lambda + \Phi \right)  \;\;\;, \\ 
{\cal L}_{II} &\equiv& 2\left( \overline{\Lambda} +
 \overline{\Phi} \right) \Gamma^M ( x_M -\rho(y_M)) 
\left( \Lambda + \Phi \right)  \;\;\;, \\
{\cal L}_{III} &\equiv& \overline{\Lambda} \Gamma^\mu x_\mu \Lambda \;\;\;.
\end{eqnarray}
It is easy to see that the action $\cal S$ can be written in terms
 of ${\cal L}_I, {\cal L}_{II},$ and  ${\cal L}_{III}$, as is seen in
 the text ( eq.~(\ref{action})).

\end{document}